\documentclass[aps,prl,twocolumn,amsfonts,amssymb,amsmath,floats,floatfix,showpacs,preprintnumbers,superscriptaddress,unsortedaddress]{revtex4}

\usepackage{graphicx}
\usepackage{dcolumn}
\usepackage{bm}

\usepackage{amsmath}

\begin{document}

\title{Elusive electron-phonon coupling in quantitative analyses of the spectral function}

\author{C.\,N. Veenstra}
\affiliation{Department of Physics {\rm {\&}} Astronomy, University of British Columbia, Vancouver, British Columbia V6T\,1Z1, Canada}
\affiliation{AMPEL, University of British Columbia, Vancouver, British Columbia V6T\,1Z4, Canada}
\author{G.\,L. Goodvin}
\affiliation{Department of Physics {\rm {\&}} Astronomy, University of British Columbia, Vancouver, British Columbia V6T\,1Z1, Canada}
\author{M. Berciu}
\affiliation{Department of Physics {\rm {\&}} Astronomy, University of British Columbia, Vancouver, British Columbia V6T\,1Z1, Canada}
\author{A. Damascelli}
\email{damascelli@physics.ubc.ca}
\affiliation{Department of Physics {\rm {\&}} Astronomy, University of British Columbia, Vancouver, British Columbia V6T\,1Z1, Canada}
\affiliation{AMPEL, University of British Columbia, Vancouver, British Columbia V6T\,1Z4, Canada}
\date{\today}

\begin{abstract}
We examine multiple techniques for extracting information from angle-resolved photoemission spectroscopy (ARPES) data, and test them against
simulated spectral functions for electron-phonon coupling. We find that, in the low-coupling regime, it is possible to extract self-energy and
bare-band parameters through a self-consistent Kramers-Kronig bare-band fitting routine
and verify the momentum independence of the self-energy along the quasiparticle dispersion.
We also show that the effective coupling parameters
deduced from the renormalization of quasiparticle mass, velocity, and spectral weight are momentum dependent and, in general, distinct from the
true microscopic coupling; the latter is thus not readily accessible in the quasiparticle dispersion revealed by ARPES.
\end{abstract}



\maketitle

Angle-resolved photoemission spectroscopy (ARPES) is a tool which, experimental difficulties aside, provides
access to the electron-removal part of the momentum-resolved spectral function $A({\bf k},\omega)$
\cite{Damascelli:reviewPS}. This quantity is an extremely rich source of information since it depends on both
the single-particle electronic dispersion $\varepsilon^b_{\bf k}$ (the so-called `bare-band') as well as the
quasiparticle self-energy $\Sigma({\bf k},\omega)\!=\!\Sigma^{\prime}({\bf
k},\omega)\!+\!i\Sigma^{\prime\prime}({\bf k},\omega)$, whose real and imaginary parts account for the energy
renormalization and lifetime of an electron in a many-body system. The single-particle spectral function is
generally written in the form:
\begin{eqnarray}
A({\bf k},\omega)=-\frac{1}{\pi}\frac{\Sigma^{\prime\prime}({\bf k},\omega)}{[\omega-\varepsilon^b_{\bf
k}-\Sigma^{\prime}({\bf k},\omega)]^2+[\Sigma^{\prime\prime}({\bf k},\omega)]^2}. \label{eqn:sf}
\end{eqnarray}
\noindent
In the case of ${\bf k}$-independent self-energies, methods based on the Lorentzian fit of momentum distribution curves [MDCs, i.e. constant
energy cuts of $A({\bf k},\omega)$] have been used to extract $\Sigma^{\prime}(\omega)$ and $\Sigma^{\prime\prime}(\omega)$ from the spectral
function measured by ARPES, and eventually to infer information on the nature and strength of the interactions dressing the quasiparticles in a
variety of complex systems \cite{Damascelli:reviewPS, vallamoly,ZX:lambda,Fink:lambda,ingleSRO,eligraphene}. However, the question of the validity of these
approaches is still a pressing one because most methods hinge on some assumption and/or approximation for the bare-band $\varepsilon^b_{\bf k}$.
More fundamental, for instance in the case of electron-boson coupling, the commonly assumed link between coupling strength and quasiparticle
renormalization, through the so-called mass enhancement factor $(1\!+\!\lambda)$, is at best merely phenomenological and its general validity
needs verification. This calls for a methodological study based on a well-behaved and momentum-independent model self-energy.

Here we investigate the possibility of extracting momentum-independent self-energies from $A({\bf k},\omega)$, without any a priori knowledge of
the bare-band $\varepsilon^b_{\bf k}$. We will test the performance of our approach on the spectral function generated with the high-order
momentum-average approximation MA$^{(n)}$ \cite{M:2006,M:2007}, for the single Holstein polaron model \cite{Holstein:original}:
momentum-independent coupling between an optical phonon and a filled one-band system of non-interacting electrons. This is a highly
oversimplified approach, in which strong interactions are not included as opposed to e.g. Ref.\,\onlinecite{werner}, and even the effect of the
Fermi sea is not accounted for (as there is no well-defined chemical potential, the latter will be positioned at the top of the first
electron-removal state). Although not applicable to correlated electron systems, the plain Holstein polaron \cite{Holstein:original} is chosen
as the minimalistic model to study the electron-phonon coupling problem. Since the MA$^{(n)}$ has been shown to be extremely accurate everywhere
in parameter space \cite{M:2007}, it will allows us to study $A({\bf k},\omega)$ over a broad range of electron-phonon coupling.

Before delving into the detailed self-energy analysis, let us illustrate the model Hamiltonian and emphasize some general aspects, relevant to
the phenomenological description of spectral functions in terms of effective coupling and renormalization parameters. Limiting ourselves to the
one dimensional case for simplicity (higher dimensions were found not to change the results qualitatively), we have used the MA$^{(n)}$
approximation to obtain self-consistent $A(k,\omega)$ with highly accurate momentum-independent $\Sigma(\omega)$, from the Holstein Hamiltonian:
\begin{eqnarray}
\mathcal{H}\!=\!\sum_{k} \varepsilon^b_k c_k^\dagger c_k\!+\!\Omega\!\sum_{Q} b_Q^\dagger b_Q\!+\!
\frac{g}{\sqrt{N}}\!\sum_{k,Q}  c_{k-Q}^\dagger c_k (b_Q^\dagger \!+\! b_{-Q}). \nonumber \\
\end{eqnarray}\label{eq:ham}
\noindent
Its terms describe, in order, an electron with dispersion $\varepsilon^b_k \!\equiv\! -2 t\, \text{cos}(k
a)$, an optical phonon with energy $\Omega$ and momentum $Q$, and the on-site electron-phonon linear coupling
[for $N$ sites with periodic boundary conditions; $c_k^\dagger$ ($c_k$) and $b_Q^\dagger$ ($b_Q$) are the
usual electron and phonon creation (annihilation) operators]. This leads to a dimensionless effective
coupling $\lambda \!\equiv\! g^2/ 2t\, \Omega $.
\begin{figure*}[t!]
\includegraphics[width=1\linewidth]{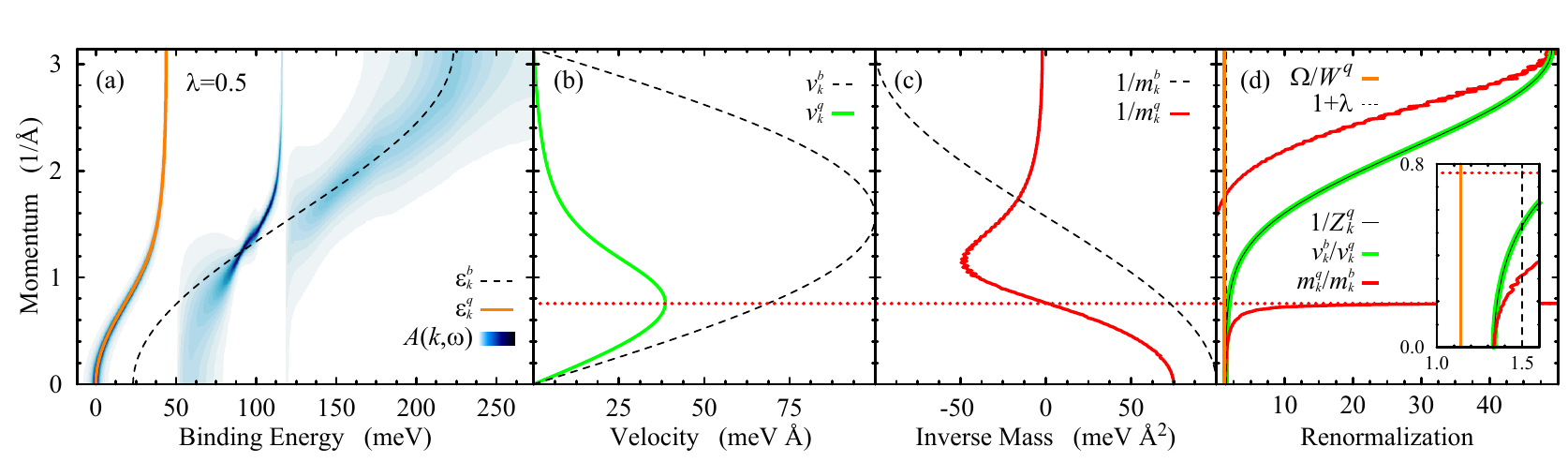}
\caption{(Color online). (a) $A(k,\omega)$ calculated within MA$^{(1)}$ for
$\Omega\!=\!50$\,meV and $\lambda \!=\!0.5$; the quasiparticle dispersion $\varepsilon^{q}_{k}$ and the
bare-band $\varepsilon^b_{k}$ are also shown. (b) Quasiparticle and bare-band velocities, $v_{k}^{q}$ and
$v_{k}^{b}$, and (c) corresponding inverse masses, $1/m_{k}^{q}$ and $1/m_{k}^{b}$,  according to the
definitions $v_k^{ }\!=\!\partial\varepsilon_{k}/\partial {k}$ and
$1/m_{k}\!=\!\partial^2\varepsilon_{k}/\partial {k}^2$. (d) Momentum-dependent quasiparticle renormalization
as obtained from $v_{k}^{b}/v_{k}^{q}$, $m_{k}^{q}/m_{k}^{b}$, and the inverse quasiparticle coherence
$1/Z^q_k$, where $Z^q_k\!=\!\int^q\! A(k,\omega)d\omega$ is the quasiparticle-only integrated spectral
weight; in the inset, these quantities are compared near $k\!=\!0$ to the renormalization factors
$\Omega/W^q$ and $(1\!+\!\lambda)$, obtained from quasiparticle bandwidth $W^q$ and dimensionless coupling
$\lambda \!=\! g^2/ 2t \, \Omega $ in our model.}\label{fig:qp_demo}
\end{figure*}
For this paper we set $a=\hbar=1$ and $t=50\text{\,meV}$, such that the bandwidth is $200\,\text{meV}$ and
the Brillouin zone is $2\pi\text{\AA}^{-1}$ wide; also note that an additional constant $1\,\text{meV}$ FWHM
Lorentzian broadening is applied, similar to an impurity scattering, to allow resolving the sharpest features
in $A(k,\omega)$.

The spectral function calculated with MA$^{(1)}$ for $\Omega\!=\!50$\,meV and $\lambda \!=\!0.5$ is presented as a false-color plot in
Fig.\,\ref{fig:qp_demo}(a); as one can see, it deviates remarkably from the one of the uncoupled bare-band $\varepsilon^b_{k}$. The
electron-removal spectrum is now comprised of a polaron quasiparticle band $\varepsilon^{q}_{k}$ (i.e., the lowest-energy bound-state of the
electron-phonon coupled system), and a continuum of excitations starting at an energy $\Omega$ below the top of the polaron band and roughly
following the original location of the bare-band $\varepsilon^b_{k}$. Based on these simulated data and on the precise knowledge of the input
value of $\lambda$, it is possible to gauge the appropriateness of estimating the electron-phonon coupling strength from the observed
renormalization of band velocities, masses, and quasiparticle coherence, as often done in the interpretation of ARPES results
\cite{Damascelli:reviewPS}. To this end, in Fig.\,\ref{fig:qp_demo}(b,c) we present the quasiparticle and bare-band velocities, $v_{k}^{q}$ and
$v_{k}^{b}$, and the inverse masses, $1/m_{k}^{q}$ and $1/m_{k}^{b}$ (see caption of Fig.\,\ref{fig:qp_demo} for definitions). The corresponding
ratios $v_{k}^{b}/v_{k}^{q}$ and $m_{k}^{q}/m_{k}^{b}$, as well as the inverse quasiparticle coherence $1/Z^q_k$ (see again caption of
Fig.\,\ref{fig:qp_demo}), are progressively larger than one the stronger the coupling strength; these quantities are usually equated to the
renormalization or mass-enhancement factor $(1\!+\!\lambda)$, providing a potential path to the quantitative estimation of the electron-phonon
coupling strength \cite{AM:1976,kulic}.

\begin{figure}[b!]
\includegraphics[width=1\linewidth]{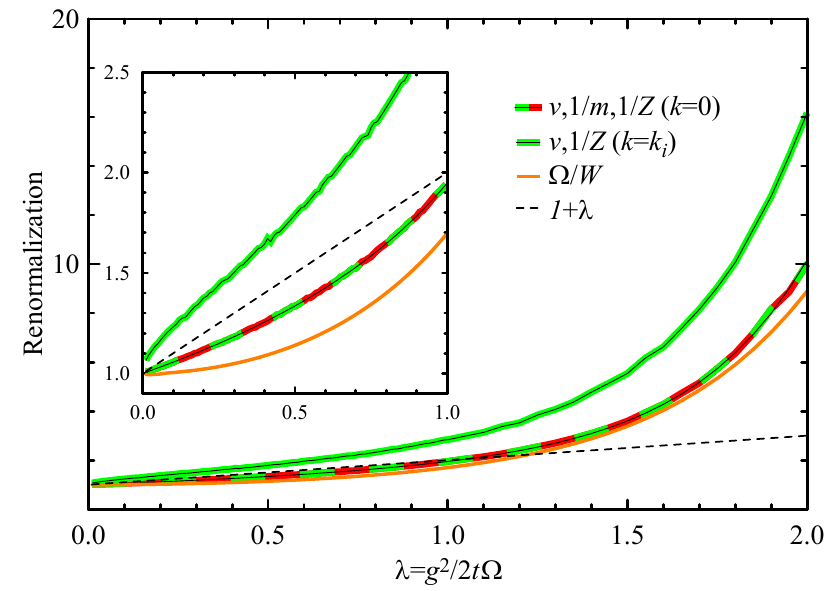}
\caption{(Color online). Renormalization parameters defined as in Fig.\,\ref{fig:qp_demo},
plotted vs. the dimensionless coupling $\lambda \!=\! g^2\!/ 2t \, \Omega$. Note that the noise in $v$ and
$1/Z$ at $k\!=\!k_i$ originates from the determination of the inflection point
$k_i$.}\label{fig:Renormalization}
\end{figure}

As shown in Fig.\,\ref{fig:qp_demo}(d), the velocity, mass, and spectral weight quasiparticle renormalizations are functions of $k$. While in
general one could expect these quantities to all be distinct, we find that $v_{k}^{b}/v_{k}^{q}\!=\!1/Z^q_k$ at all $k$, which is a direct
consequence of the $k$-independence of the self-energy derived from the Holstein Hamiltonian: by Taylor expanding $\Sigma(k,\omega)$ in the
vicinity of the quasiparticle pole, i.e. $\omega\!=\!\varepsilon^{q}_{k}\!+\!\delta\omega$, one can approximate the Green's function as
$G(k,\omega)\!\backsimeq\!Z_k/(\omega\!-\!\varepsilon^{q}_{k})$ in terms of the quasiparticle coherence $Z_k^q\!=\!1/[1-\partial
\Sigma(k,\omega)/\partial \omega \mid_{\omega\!=\!\varepsilon^{q}_{k}}]$, and obtain for the quasiparticle velocity
$v_k^q\!=\!Z_k^q[v_k^b+\partial \Sigma(k,\omega)/\partial k \mid_{\omega\!=\!\varepsilon^{q}_{k}}]$; the latter reduces to $v_k^q\!=\!Z_k^q
v_k^b$ if $\partial \Sigma/\partial k\!=\!0$. For the mass renormalization, we obtain $m_{k}^{q}/m_{k}^{b}\!=\!1/Z^q_k$ only for $k\!=\!0$ and
$\pi$, which is simply a consequence of the fact that at the band extrema the velocities are zero and the corresponding rate of change away from
the extrema has to follow the acceleration [see Fig.\,\ref{fig:qp_demo}(d) and its inset; the horizontal dashed line emphasizes the divergence
of $m_{k}^{q}/m_{k}^{b}$ due to the vanishing of $1/m_{k}^{q}$]. Most importantly, $v_{k}^{b}/v_{k}^{q}$, $m_{k}^{q}/m_{k}^{b}$, and $1/Z^q_k$
cannot be compared directly to the momentum-independent renormalization factors $\Omega/W^q$ and $(1\!+\!\lambda)$, obtained from the
quasiparticle bandwidth $W^q$ and the dimensionless coupling $\lambda \!=\! g^2/ 2t \, \Omega $ in our model [Fig.\,\ref{fig:qp_demo}(d),
inset].
Although coincidences in values can be observed at some electron momenta, these are purely circumstantial and
cannot be generalized.

The results in Fig.\,\ref{fig:qp_demo} demonstrate that in general one cannot directly extract a quantitative value for the coupling strength
$\lambda$ from the observed renormalization of quasiparticle velocity, mass, and coherence (and this without even considering further
complications originating from the electron-correlation driven renormalization of electron-phonon coupling in real materials
\cite{manganites,ingleSRO}). If one was to do that, the extracted value would vary with the chosen observable and the specific $k$ value, making
any conclusion arbitrary (e.g., at $k\!=\!\pi$ these renormalizations overestimate $\lambda$ by a factor of almost 100). One may wonder,
however, if any of these quantities scaled as $(1\!+\!\lambda)$ upon increasing $\lambda$, so that if not the exact value at least the trend of
the coupling strength could be captured, for example in an experiment performed as a function of doping. In Fig.\,\ref{fig:Renormalization} we
follow each of these quantities and the quasiparticle bandwidth as a function of $\lambda$, contrasted against $(1\!+\!\lambda)$. For momentum
dependent quantities we must choose a $k$ value: we plot $v_{0}^{b}/v_{0}^{q}$, $m_{0}^{q}/m_{0}^{b}$, and $1/Z^q_0$ at $k=0$ where they all
coincide, as well as $v_{k_i}^{b}/v_{k_i}^{q}$ and $1/Z^q_{k_i}$ at the inflection point $k\!=\!k_i$ of the quasiparticle band
$\varepsilon^{q}_{k}$ where $m_{k_i}^{q}/m_{k_i}^{b}$ diverges. All quantities deviate dramatically from $(1\!+\!\lambda)$ at large coupling
values (e.g., overestimating the microscopic coupling by a factor of 4 to 8 at $\lambda\!=\!2$), and are also rather poor indicators of the
coupling strength in the low-coupling regime ($\lambda\!<\!1$; Fig.\,\ref{fig:Renormalization} inset). Interestingly, the inflection point
velocity renormalization $v_{k_i}^{b}/v_{k_i}^{q}$ is linear in $\lambda$ over a range that scales with the ratio $\Omega/t$. However, even this
term deviates from $(1\!+\!\lambda)$ in a way dependent on the details of the model, such as the shape of the bare-band $\varepsilon^b_{k}$; in
general, it cannot be used for the quantitative estimate of $\lambda$.

Determining $\varepsilon^b_{k}$ is a key step also in the attempt of extracting real and imaginary parts of the self-energy from $A(k,\omega)$,
which will be the focus of the remainder of the paper. We analyze $A(k,\omega)$ in terms of MDCs at constant energy $\omega=\tilde\omega$. Since
the self-energy in the present model is $k$-independent, i.e. $\Sigma^{\prime}_{\tilde\omega}$ and $\Sigma^{\prime\prime}_{\tilde\omega}$ are
constant, as long as $\varepsilon^b_{k}$ can be linearized in the vicinity of the MDC peak maximum at $k=k_m$, the MDC lineshape is Lorentzian.
Note that the converse is not true: a Lorentzian MDC lineshape is not a sufficient condition to conclude $\Sigma\!=\!\Sigma(\omega)$
\cite{Randeria:DidKDep}, although the overlap of $v_{k}^{b}/v_{k}^{q}$ and $1/Z^q_{k}$ (Fig.\,\ref{fig:qp_demo}) confirms the $k$-independence
along the quasiparticle dispersion.
\begin{figure}[t!]
\includegraphics[width=1\linewidth]{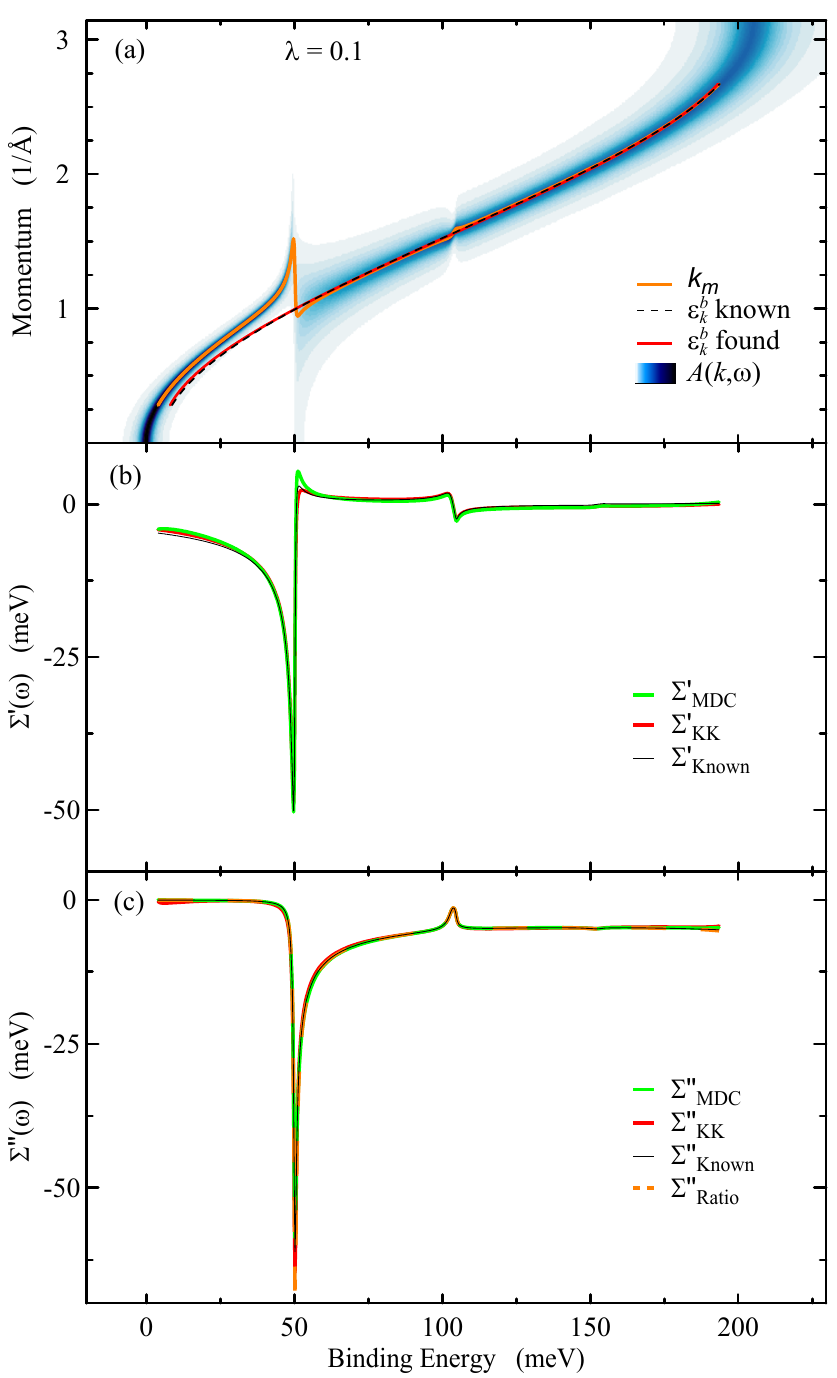}
\caption{(Color online). (a) $A(k,\omega)$ calculated within MA$^{(1)}$ for
$\Omega\!=\!50$\,meV and $\lambda \!=\!0.1$; also shown are the $k_m$ path of MDC maxima, as well as the
known bare-band and the one found through the KKBF analysis. (b,c) Real and imaginary part of the self-energy
from the model ($\Sigma_{known}$), the bare-band and MDC fitting routine ($\Sigma_{MDC}$), and the KK
transform of $\Sigma^{\prime\prime}_{MDC}$ ($\Sigma^{\prime}_{KK}$) and $\Sigma^{\prime}_{MDC}$
($\Sigma^{\prime\prime}_{KK}$). In (c) the MDC ratio results, $\Sigma^{\prime\prime}_{ratio}\!=\!-\Delta
k_m/A_{0}$, are also shown.}\label{fig:kkbf_results}
\end{figure}
By Taylor expanding $\varepsilon^b_{k}$ about the MDC peak maximum at $k=k_m$, i.e.
$\varepsilon^b_{k}\!=\!\varepsilon^b_{k_m}\!+\!v^b_{k_m}\!\cdot (k-k_m)\!+\!...$, and noticing that
$\tilde\omega\!=\!\varepsilon^b_{k_m}\!+\!\Sigma^{\prime}_{\tilde\omega}$, we can rewrite Eq.\,\ref{eqn:sf}
as:
\begin{equation}
A_{\tilde\omega}(k)\!\backsimeq\! \frac{A_0}{\pi} \frac{ \Delta k_m } { (k-k_m)^2 + (\Delta k_m)^2 },
\label{eqn:se_lorentzian}
\end{equation}
where $\Delta k_m \!=\! - \Sigma^{\prime\prime}_{\tilde\omega}/v^b_{k_m}$ is the half-width half-maximum of
the Lorentzian MDC and $A_0 \!=\! 1/v^b_{k_m}\!=\!\int\!A_{\tilde\omega}(k)dk$. If the bare-band is not known
it is possible to fit it, within an arbitrary offset, to any functional form which provides a value and
derivative using a Kramers-Kronig bare-band fitting (KKBF) routine. This is done by first tracking $k_m$ and
$\Delta k_m$ for every $\tilde\omega$ through a Lorentzian fit of the MDC (Eq.~\ref{eqn:se_lorentzian}), and
then choosing $\varepsilon^b_{k}$ parameters such that
$\Sigma^{\prime}_{MDC}\!\equiv\tilde\omega\!-\!\varepsilon^b_{k_m}$ and
$\Sigma^{\prime\prime}_{MDC}\!\equiv\!- v^b_{k_m} \Delta k_m$ are self-consistent with $\Sigma^{\prime}_{KK}$
and $\Sigma^{\prime\prime}_{KK}$ calculated from the Kramers-Kronig (KK) relations:
\begin{equation}
\Sigma^{\prime\,,\,\prime\prime}_{KK}(k,\omega) = \pm \frac{1}{\pi} \mathcal{P} \int^\infty_{-\infty}
\partial \xi
\frac{\Sigma^{\prime\prime\,,\,\prime}_{MDC}(k,\xi)} {\xi-\omega}. \label{eqn:kk}
\end{equation}
\noindent
In our implementation (Fig.\,\ref{fig:kkbf_results}), a simple third order polynomial was used with an initial guess found by fitting MDC peak
maxima.  We then used the Levenberg-Marquardt Algorithm as implemented in the mpfit package for IDL to vary band parameters.  We found that the
standard sum-of-squares minimization did not perform as well as a concave-down function, since it placed too much weight on outlying points far
away. In order to evaluate the integrals in Eq.\,\ref{eqn:kk} with a finite region of data, biased inverse polynomial fits where used to
extrapolate tails before a Fourier-based transform was performed.

The method outlined here varies slightly from techniques previously described in the literature, which often deal with data very close to the
Fermi energy and generally reduce the possible functional forms for $\varepsilon^b_{k}$ substantially \cite{Kordyuk:2005,Kaminski:2005}. While
these methods present an exact solution for $A(k,\omega)$ based on a reduced bare-band, the present approach imposes no restrictions on
$\varepsilon^b_{k}$ (other than it be differentiable near $k_m$), allowing fitting based on a wider variety of bare-band models and using KKBF
to vary $\varepsilon^b_{k}$ parameters. The results of the KKBF procedure are presented for $\lambda\!=\!0.1$ in Fig.\,\ref{fig:kkbf_results}
and show that, in the low-coupling regime, the ${\it found}$ bare-band and self-energies ($\Sigma_{MDC}$) agree well with the ${\it known}$
quantities from our model (as long as the MDC analysis is applicable: $k_{m}$ is far from band extrema where velocities vanish, and MDC widths
are suitably small that the bare-band may be approximated as linear around $k_{m}$). Note that the self-energies are evaluated along the
$(k_m,\tilde\omega)$-path of the MDC maxima, along which the analysis is performed; this path deviates significantly from either
$\varepsilon^q_{k}$ or $\varepsilon^b_{k}$ close to the sharper one- and two-phonon structures [Fig.\,\ref{fig:kkbf_results}(a)].
\begin{figure}[t!]
\includegraphics[width=1\linewidth]{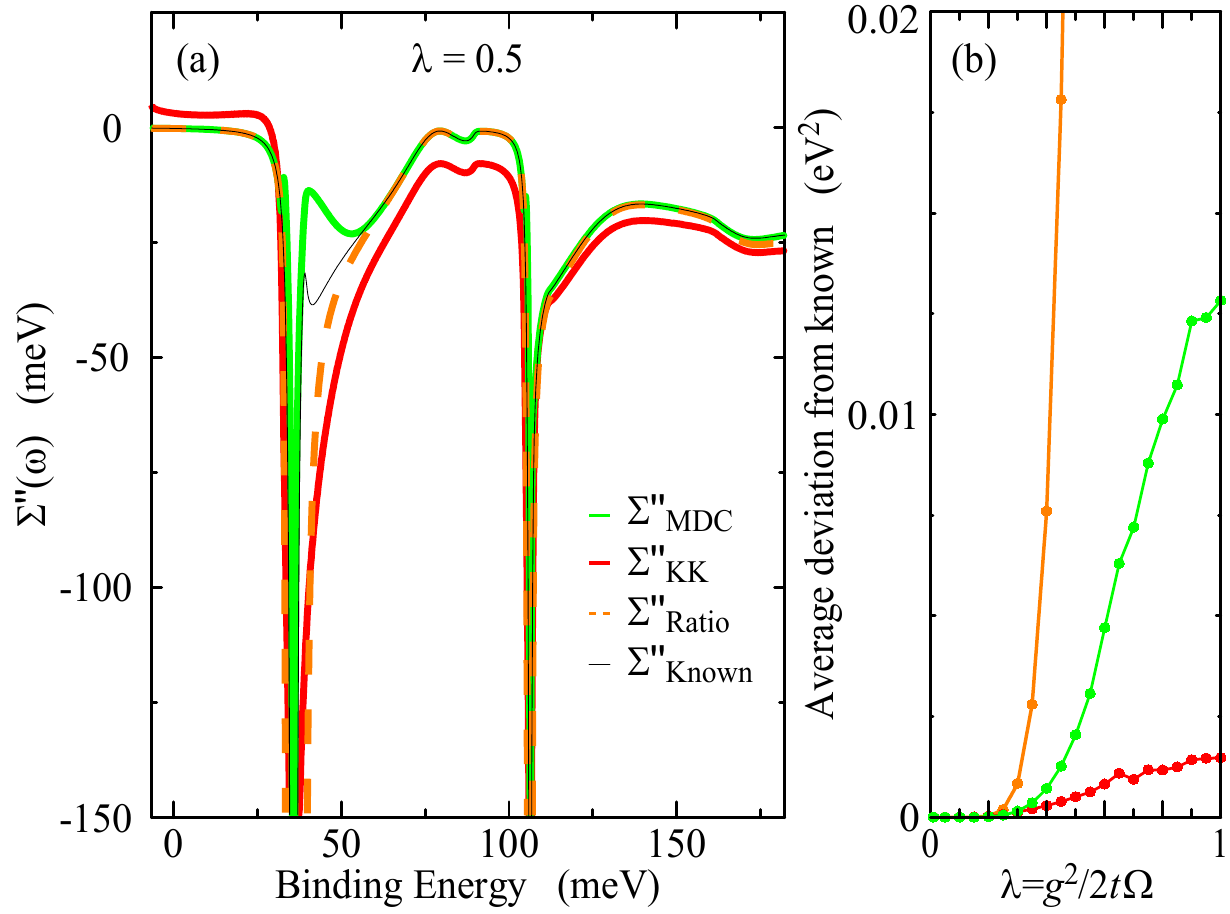}
\caption{(Color online). (a) Various estimates for the imaginary part of the self-energy,
defined as in the caption of Fig.\,\ref{fig:kkbf_results}, for $A(k,\omega)$ calculated within MA$^{(1)}$
\,for\, $\Omega\!=\!50$\,meV and $\lambda \!=\!0.5$.\,(b) Deviation (i.e., average of the squared difference
at each $k_m$) between estimated and known self-energies vs. $\lambda$.}\label{fig:kkbf_failure}
\end{figure}
The reliability of the results is confirmed by the agreement between $\Sigma_{MDC}$ and $\Sigma_{KK}$ over
the whole range, which provides an internal self-consistency check.

Although very satisfactory at low coupling, at larger coupling the results from the KKBF procedure become
progressively less accurate. This can be seen already for $\lambda\!=\!0.5$ in
Fig.\,\ref{fig:kkbf_failure}(a), where $\Sigma^{\prime\prime}_{MDC}$ and $\Sigma^{\prime\prime}_{KK}$ fail to
reproduce $\Sigma^{\prime\prime}_{known}$ in the proximity of the sharp phonon-induced structures
($\Sigma^{\prime\prime}_{KK}$ has also picked up different offsets in the different flatter parts of the
spectrum, making setting its overall offset difficult). For a more comprehensive description of the deviation
of these quantities from $\Sigma^{\prime\prime}_{known}$ upon increasing $\lambda$, in
Fig.\,\ref{fig:kkbf_failure}(b) we present the average of the squared difference at each $k_m$ between the
estimated and known self-energies versus $\lambda$: for both $\Sigma^{\prime\prime}_{MDC}$ and
$\Sigma^{\prime\prime}_{KK}$, this estimate shows a rapid and consistent increase with $\lambda$, indicating
a failure of the KKBF analysis and similar methods in the intermediate to strong electron-phonon coupling
regime.

Before concluding, we will point out an alternative, possibly more practical approach, which allows us to tackle the problem over a larger range
of $\lambda$, at least for $\Sigma^{\prime\prime}$. In our method, determining real and imaginary parts of the self-energy hinges on finding a
proper expression for the bare-band through the KKBF routine. The imaginary part however,
$\Sigma^{\prime\prime}_{\tilde\omega}\!=\!-v^b_{k_m}\Delta k_m$, only requires knowledge of $v^b_{k_m}$, and this can be obtained directly from
$A(k,\omega)$ in two independent ways. The first one is through the momentum integral of $A_{\tilde\omega}(k)$ in Eq.\,\ref{eqn:se_lorentzian},
which returns directly $A_0\!=\!1/v_{k_m}^b$; this allows a simple estimate of $\Sigma^{\prime\prime}$ [and equally accurate to KKBF's for
$\lambda\!=\!0.1$, Fig.\,\ref{fig:kkbf_results}(c)] from the MDC width/integral ratio: $\Sigma^{\prime\prime}_{ratio}\!=\!-\Delta k_m/A_{0}$.
The second one is through the equivalence $v_{k}^{b}/v_{k}^{q}\!=\!1/Z^q_k$ discussed in the context of Fig.\,\ref{fig:qp_demo}(d), and the
possibility of estimating $v_{k}^{q}$ and $Z^q_k$ directly from the data as the momentum derivative and energy integral of the quasiparticle
band $\varepsilon^{q}_{k}$ (Fig.\,\ref{fig:qp_demo}): $\Sigma^{\prime\prime}_{MDC}\!=\!-v_{k}^{q}\Delta k_m/Z^q_k$. The so-obtained
$\Sigma^{\prime\prime}_{ratio}$ and $\Sigma^{\prime\prime}_{MDC}$ are compared in Fig.\,\ref{fig:kkbf_failure}(a); although especially
$\Sigma^{\prime\prime}_{ratio}$ deviates strongly from $\Sigma^{\prime\prime}_{known}$ upon increasing $\lambda$ at the sharp phonon-induced
features [Fig.\,\ref{fig:kkbf_failure}(b)], the general behavior is that when
$\Sigma^{\prime\prime}_{ratio}\!\approx\!\Sigma^{\prime\prime}_{MDC}$ they also match $\Sigma^{\prime\prime}_{known}$ almost exactly. Thus,
$\Sigma^{\prime\prime}_{ratio}$ and $\Sigma^{\prime\prime}_{MDC}$ can be used to obtain a precise anchor mesh for $\Sigma^{\prime\prime}$ over a
range of energies, without the critical step of finding the bare-band through KKBF.

In summary we have shown that, at variance with a common phenomenological practice in the interpretation of
ARPES data, even in the simplest case of electron-phonon coupling described by the Holstein model it is not
possible to obtain the microscopic coupling strength from the observed renormalization of quasiparticle
coherence, velocity, mass, or bandwidth (e.g., $v_{k_i}^{b}/v_{k_i}^{q}$ would overestimate $\lambda$ by a
factor of ~2, 3, and 8 for $\lambda\!=\!1$, 1.5, and 2). In this sense, the coupling $\lambda$ still remains
an elusive quantity. Through the KKBF analysis we can however gain access to bare-band and self-energies,
which if properly modelled might provide information on the nature and strength of the underlying
interactions, at least in the momentum independent, low-coupling regime.

We gratefully acknowledge S. Johnston, T.P. Devereaux, F. Marsiglio, and G.A. Sawatzky for many useful discussions. This work was supported by
the Killam Program (A.D.), the Alfred P. Sloan Foundation (M.B. and A.D.), CRC Program (A.D.), NSERC, CFI, CIFAR Quantum Materials and
Nanoelectronics Programs, and BCSI.


\bibliography{KKBF_resub}

\end{document}